\documentclass{optica-article}
\journal{opticajournal} 
\articletype{Research Article}

\usepackage{color}

\usepackage{xspace}

\newcommand{\invitro}{\textit{in vitro}\xspace}

\newcommand{\enface}{\textit{en face}\xspace}

\newcommand{\um}{\(\mu\)m\xspace}
\newcommand{\umcubic}{\(\mu\)m$^3$\xspace}
\newcommand{\uminvcubic}{\(\mu\)m$^{-3}$\xspace}
\newcommand{\pixcubic}{\ensuremath{\mathrm{pix}^3}\xspace}

\newcommand{\ift}[1]{\mathcal{F}^{-1}\left[{#1}\right]\xspace}
\newcommand{\Nsh}{$N_{\mathrm{sh}}$\xspace}
\newcommand{\NRIN}{$N_{\mathrm{RIN}}$\xspace}
\newcommand{\Nno}{$N_{\mathrm{no}}$\xspace}
\newcommand{\etal}{\textit{et al.\@}\xspace}

\graphicspath{{.}{./Figures}}

\begin{document}

\title{Optical-coherence-tomography-based deep-learning scatterer-density estimator using physically accurate noise model}

\author{Thitiya Seesan,\authormark{1}
	Pradipta Mukherjee,\authormark{1}
	Ibrahim Abd El-Sadek,\authormark{1,2}
	Yiheng Lim,\authormark{1}
	Lida Zhu,\authormark{1}
	Shuichi Makita,\authormark{1}
	and Yoshiaki Yasuno\authormark{1,*}}

\address{\authormark{1}Computational Optics Group, University of Tsukuba, Tsukuba, Ibaraki 305-8573, Japan\\
	\authormark{2}Department of Physics, Faculty of Science, Damietta University, New Damietta City 34517, Damietta, Egypt}

\email{\authormark{*}yoshiaki.yasuno@cog-labs.org} 
\homepage{https://cog-news.blogspot.com/}


\begin{abstract*} 
We demonstrate a deep-learning-based scatterer density estimator (SDE) that processes local speckle patterns of optical coherence tomography (OCT) images and estimates the scatterer density behind each speckle pattern.
The SDE is trained using large quantities of numerically simulated OCT images and their associated scatterer densities.
The numerical simulation uses a noise model that incorporates the spatial properties of three types of noise, i.e., shot noise, relative-intensity noise, and non-optical noise.
The SDE's performance was evaluated numerically and experimentally using two types of scattering phantom and \invitro tumor spheroids.
The results confirmed that the SDE estimates scatterer densities accurately.
The estimation accuracy improved significantly when compared with our previous deep-learning-based SDE, which was trained using numerical speckle patterns generated from a noise model that did not account for the spatial properties of noise. 
\end{abstract*}

\section*{Publication information}
The final version of this manuscript has been published in Biomedical Optics Express and is available as: \\
Thitiya Seesan, Pradipta Mukherjee, Ibrahim Abd El-Sadek, Yiheng Lim, Lida Zhu, Shuichi Makita, and Yoshiaki Yasuno, ``Optical-coherence-tomography-based deep-learning scatterer-density estimator using physically accurate noise model,'' \textit{Biomed. Opt. Express} \textbf{15}, 2832-2848 (2024). \url{https://doi.org/10.1364/BOE.519743}.

\section{Introduction}
\label{sec:intro}

Optical coherence tomography (OCT) is a non-invasive imaging modality that is used to provide high-resolution structural information about biological tissues\cite{huang_optical_1991,fujimoto_optical_2000,fujimoto_optical_2003}.
The anatomical images provided by OCT are used widely in clinical diagnosis \cite{Chen2007}.

In addition to anatomical investigations, OCT-based assessments of the optical properties of tissue have also been studied and have provided useful biomarkers.
One commonly investigated optical property of biological tissues is the attenuation coefficient (AC) \cite{liu_tissue_2017, chang_review_2019}, and the AC is considered to be related to the tissue density.
AC measurements are useful in a wide variety of applications, including investigation of tumor spheroid necrosis \cite{huang_optical_2017, huang_longitudinal_2020} and distinguishing between normal and cancerous tissues \cite{cauberg_quantitative_2010,zhao_ex_2012, muller_prostate_2016}.
However, AC measurements are strongly influenced by the measurement configuration and conditions, which including the system confocality, the depth position of the focus, and the presence of aberrations \cite{van_leeuwen_measurement_2003,stefan_determination_2018,li_robust_2020}, and these factors can limit the accuracy and reliability of the AC measurements.
Although several methods have been proposed to compensate for these effects, they generally require hard-wired assumptions to be made \cite{van_leeuwen_measurement_2003} or multiple measurements to be performed \cite{baumann_signal_2019}. 

Rather than use the AC to assess biological tissue density, we have proposed a deep-learning based method that estimates the scatterer density of the tissue directly \cite{seesan_intensity-invariant_2020, seesan_sample_2021, seesan2022}.
In this work, we denote this method as the scatterer density estimator (SDE).  
The SDE analyzes the local spatial patterns of an OCT intensity image, i.e., the speckle pattern, using a convolutional neural network (CNN) and then estimates the scatterer density. 
The CNN was trained using fully numerically simulated OCT speckle patterns that were generated by a simple scalar-optics-based OCT simulator.
This approach provides significant amounts of training data, and the data sets reflect the variety of the parameters involved in OCT imaging, including the resolution, the signal-to-noise ratio (SNR), and the sample scatterer density.

In our previous study\cite{seesan2022}, we noted that the CNN can estimate two primary quantities: the speckle contrast and the resolution. 
Hillman \etal showed that the local speckle contrast of OCT has a monotonic and negative relationship with the number of scatterers contained with in a three-dimensional (3D) resolution volume, which is designated the effective number of scatterers (ENS) \cite{hillman_correlation_2006}. 
In addition, Kurokawa \etal demonstrated that the axial and lateral resolutions of OCT can be estimated based on a local speckle pattern \cite{kurokawa_-plane_2015}.
Therefore, we have anticipated that the CNN will primarily estimate the speckle contrast and the resolutions, and have then estimated the scatterer density based on these quantities. 

This previous SDE has provided promising results, but its scatterer density estimation accuracy was limited when it was applied to real experimental data, particularly if the scatterer density or the SNR is low \cite{seesan2022}. 
By considering the possible mechanism for the estimator described above, we hypothesized that the low estimation accuracy for both the speckle contrasts and the resolutions may be limiting factors for the scatterer density.
In addition, the estimation accuracy of the scatterer density falls when the SNR is low, i.e., in cases in which the noise forms a significant component of the OCT image.
As a result, we hypothesized that our simulation-generated training data (i.e., the speckle patterns) did not actually reflect the physical properties of the noise accurately.
This inaccuracy in the noise modeling process may have then caused the inaccuracies in both the speckle contrast and the resolution estimations.

In this paper, we introduce a new physically realistic noise model and demonstrate improved accuracy in the SDE.
This new noise model incorporates the spatial properties of the OCT noises including shot noise, relative intensity noise (RIN), and non-optical noise.
The accuracy of the SDE is then validated using numerically generated OCT images, along with OCT images of two types of scattering phantoms and \invitro tumor spheroids.
Our results demonstrate significant improvement in the scatterer density estimation accuracy, particularly under low SNR conditions.

\section{Principle}
\label{sec:principle}

\subsection{Neural network-based scatterer density estimator}
\label{sec:conceptSDE}
Our CNN-based SDE estimates the scatterer density from a small spatial pattern of an OCT image, i.e., from a local speckle pattern with an image size of 16 $\times$ 16 pixels.
To train the CNN model, we used numerical speckle patterns that were generated by a simple scalar-optics-based OCT simulator, which is designated a ``speckle generator.''
The speckle generator generates a speckle pattern from an arbitrary parameter set that consists of the scatterer density, the axial and lateral resolutions, and the SNR.
The speckle pattern is generated by convolving a 3D scatterer distribution map with a 3D complex point spread function and subsequently adding complex Gaussian noise.
The details of the speckle generator are presented in Section \ref{sec:speckleGenerator}.

In our previous study, the noise was assumed to be fully spatially decorrelated.
In other words, the noise signals at the individual pixels are assumed to be fully independent of each other.
In the new SDE, we introduce a more physically accurate noise model that accounts for the spatial correlation properties of the noise, as will be described in the next section.

\subsection{Noise Model}
\label{sec:noise_model}
Similar to our previous noise model, the new noise model has a complex Gaussian distribution, i.e., both the real and imaginary parts of the noise are normally distributed with a zero mean. 
However, our new noise model accounts for the different spatial correlation properties of the three types of OCT noise, comprising shot noise, RIN, and non-optical noise.
These three noise types have different spatial extents, and these differences originated from the different dependencies of these noise types on the light source spectrum.

As elucidated in prior works \cite{Leitgeb2003OpEx, deBoer2003OL}, the energies of shot noise and RIN are proportional to light intensity and the square of light intensity, respectively. 
Consequently, the amplitudes of shot noise and RIN are proportional to the square root and the first-root of light intensity, respectively. 
Note that the first root of light intensity is the light intensity itself. 
Conversely, both the energy and amplitude of non-optical noise remain invariant with respect to changes in light intensity. 
Thus, the amplitudes of shot noise and RIN are proportional to the square root and the first-root of the light intensity, respectively. 
Note that the first-root of the light power is the light power itself. 
On the other hand, both the energy and amplitude of non-optical noise are independent of light intensity. 
And hence, the amplitudes of the shot noise and the RIN are proportional to the square-root and the first-root of the light intensity, respectively. 
Note that the first-root of the light power is the light power itself. 
On the other hand, both the energy and amplitude of non-optical noise is independent from the light intensity. 
And hence, when the light source spectral intensity is denoted by $S(k)$, the noise within the wavenumber domain can be described as
\begin{equation}
	\label{eq:Na_kDomain}
	N_a(k) = c_1 \sqrt{S(k)}N_{\mathrm{sh}}(k) + c_2 S(k) N_{\mathrm{RIN}}(k) + c_3 N_{\mathrm{no}}(k),
\end{equation}
where $k$ is the wavenumber, and \Nsh, \NRIN, and \Nno are normally distributed random variables that correspond to the shot noise, the RIN, and the non-optical noise, respectively.
All these variables have a zero mean and the same standard deviations, but they are independent of each other.
In addition, the constants $c1$, $c2$, and $c3$ are proportionality constants.

As the equation shows, the amplitude expectation of the shot noise (i.e., the first term on the right-hand side) is a function of $k$, and is proportional to the square root of the spectral intensity at each value of $k$.
This is because the shot noise energy, which is the squared power of the amplitude, is proportional to the spectral intensity.
Similarly, the amplitude expectation of the RIN (i.e., the second term) is also a function of $k$, and is also proportional to the spectral intensity.
This corresponds to the fact that the RIN energy is proportional to the squared power of the spectral intensity.
In contrast, the non-optical noise (i.e., the third term) is independent of the light source spectrum and its amplitude expectation remains constant over the range of $k$.

The spectral-intensity dependence of the noise causes a spatial correlation of the noise along the depth direction.
The noise in the depth domain can be determined by taking the Fourier transform of Eq. (\ref{eq:Na_kDomain}) as follows
\begin{equation}
	\label{eq:Na_zDomain}
	\ift{N_a (z)} = c_1 \ift{\sqrt{S(k)}}(z) * N_{\mathrm{sh}}'(z) + c_2 \ift{S(k)}(z) * N_{\mathrm{RIN}}'(z) + c_3 N_{\mathrm{no}}'(z)
\end{equation}
where $z$ is the depth and is a Fourier pair of $k$.
$\ift{\quad}$ denotes an inverse Fourier transform and $N_{\mathrm{sh}}'(z)$, $ N_{\mathrm{RIN}}'(z)$, and $ N_{\mathrm{no}}'(z)$ are the Fourier transforms of $N_{\mathrm{sh}}(k)$, $ N_{\mathrm{RIN}}(k)$, and $ N_{\mathrm{no}}(k)$, respectively.
Because $N_{\mathrm{sh}}(k)$, $ N_{\mathrm{RIN}}(k)$, and $ N_{\mathrm{no}}(k)$ are all normally distributed Gaussian noise, $N_{\mathrm{sh}}'(z)$, $ N_{\mathrm{RIN}}'(z)$, and $ N_{\mathrm{no}}'(z)$ become normally distributed complex Gaussian noise.
As the equation shows, the shot noise is convolved with the Fourier transform of $\sqrt{S(k)}$, and the result is the spatial correlation of the shot noise along the depth direction. 
Similarly, the RIN is convolved with the Fourier transform of $S(k)$ and this operation also results in a spatial correlation along the depth direction. 

Notably, because $\sqrt{S(k)}$ can be wider than $S(k)$, $\ift{\sqrt{S(k)}}(z)$ can then be narrower than $\ift{S(k)}(z)$.
This suggests that the spatial correlation distance of the shot noise is shorter than that for the RIN.
In contrast to the shot noise and the RIN, i.e., the optical noise types, the non-optical noise (the third term on the right-hand side) shows no spatial correlation.
The spatial correlation distances for the three types of noise contained in the new noise model and in the old noise model used in our previous SDE \cite{seesan2022} are summarized in Table \ref{tab:spatialCorrelation}, in addition to the correlation distance of the OCT signal.
\begin{table}
	\caption{
		Summary of spatial correlation distance characteristics of three noise types and the OCT signal.
		In the new noise model, each type of noise has different spatial correlation properties, whereas these properties \cite{seesan2022} are identical for all noise types in the old noise model.
	}
	\label{tab:spatialCorrelation}
	\centering\includegraphics[width=7cm]{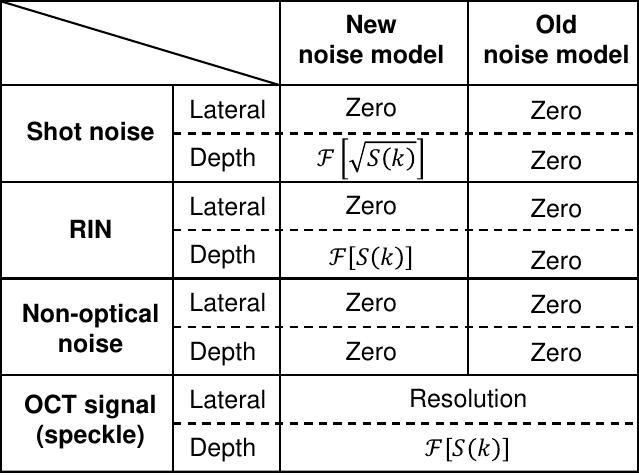}
\end{table}

In this study, we consider scanning OCT, rather than a full-field OCT.
Because each A-line of the scanning OCT is acquired at a different time, none of the three noise types have spatial correlation along the lateral direction.
We expect that the different spatial properties of the noises and the OCT signal will be used by the CNN to estimate the speckle contrast and the resolutions accurately.
This will then enable more accurate estimation of the scatterer density than our old CNN model, as will be shown in Section \ref{sec:results}.

\section{Implementation and validation method}
\subsection{Speckle generation and CNN-model training}
\label{sec:speckleGenerator}
\begin{figure}
	\centering
	\includegraphics[width=3in]{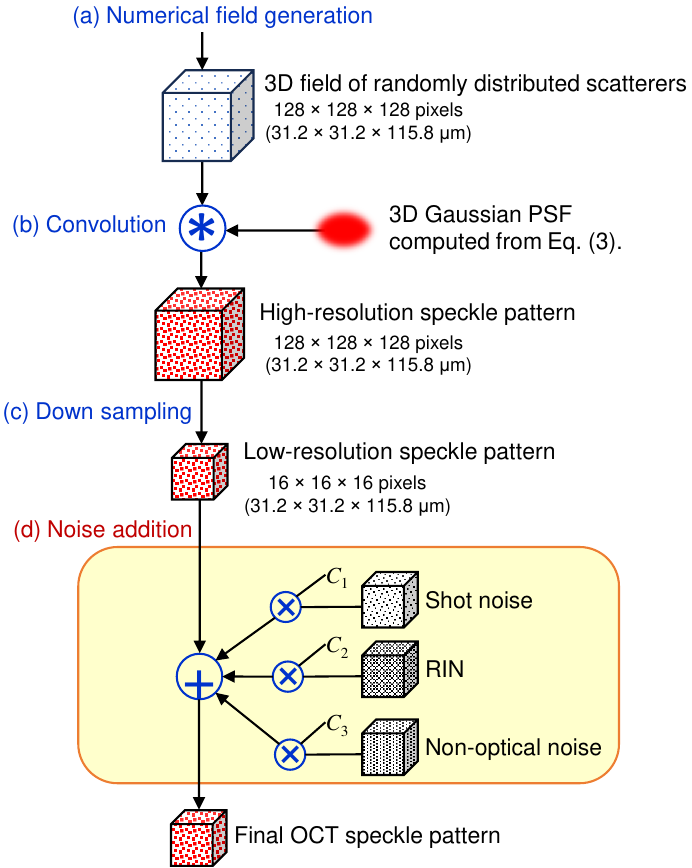}
	\caption{%
		The flow diagram illustrates the OCT speckle generation process. 
		Initially, a 3D random scattering field is generated (a), which is then convolved with a 3D Gaussian PSF that is computed from Eq.\@ (\ref{eq:PSF}). The resulting high-resolution speckle pattern is downsampled (c). Subsequently, three types of noise, i.e., shot noise, RIN, and non-optical noise, are added to obtain the final OCT speckle pattern which is used for training the CNN.
	}
	\label{fig:speckeGenerationFlow}
\end{figure}
The first step in the CNN model training process is the generation of the OCT speckle patterns, i.e., the training datasets.
This step is identical to that described in Section 2.1 of \cite{seesan2022}, with the exception of the noise model.
Because the details of this numerical speckle pattern generation process have been described elsewhere, we describe the process only briefly here.
In this process, we first generate the 3D numerical fields using randomly distributed scatterers [(a) in Fig.\@ \ref{fig:speckeGenerationFlow}].
Here, the field has a size of 128 $\times$ 128 $\times$ 128 pixels (31.2 \um $\times$ 31.2 \um for the lateral directions and 115.8 \um for the axial direction), and the pixels with a scatterer have an amplitude of unity but a random phase.  
This random phase represents the sub-wavelength depth position of the scatterer.
The pixels without a scatterer have zero values.
The 3D field of scatterers is then convolved with a 3D complex point spread function of the OCT [(b) in Fig.\@ \ref{fig:speckeGenerationFlow}], which is assumed to have a 3D Gaussian distribution computed from the following equation:
\begin{equation}
	\label{eq:PSF}
	\mathrm{PSF}(x,y,z) \propto \exp{- \left[
		4\left(\frac{x}{\Delta x}\right)^2
		+ 4\left(\frac{y}{\Delta y}\right)^2
		+ 2\ln{2}\left(\frac{z}{\Delta z}\right)^2
		\right]},
\end{equation} 
where $\Delta x$ and $\Delta y$ are the lateral resolutions defined as the $1/e^2$-width of the squared amplitude of PSF, and $\Delta z$ is the axial resolution defined as the full-width-half-maximum (FWHM) of the squared amplitude.
The two lateral resolutions are assumed to be identical.
The resolutions are randomly selected as described in the next paragraph. 
The 3D numerical field is then down-sampled from 128 $\times$ 128 $\times$ 128 pixels to 16 $\times$ 16 $\times$ 16 pixels, in keeping with the original physical field size [(c) in Fig.\@ \ref{fig:speckeGenerationFlow}].
The pixel size after down-sampling is then 1.95 \um for the lateral directions and 7.24 \um for the axial direction.
After down-sampling of the field, we added the shot noise, the RIN, and the non-optical noise which follow the noise model described in Section \ref{sec:noise_model} [(d) in Fig.\@ \ref{fig:speckeGenerationFlow}].

To train the CNN model, we generated 80,000 3D speckle patterns and then extracted 1,280,000 2D cross-sectional speckle patterns with dimensions of 16 $\times$ 16 pixels.
Each 3D speckle pattern was generated with different and randomly selected values for the scatterer density, the resolutions, and the noise energies of the shot noise, the RIN, and the non-optical noise.
The axial and lateral resolutions here are independent of each other and range from 3 to 30 \um.
The noise energies for each type of noise are randomly selected as the SNR with respect to each noise type in the range from 0 to 100 dB, where 0 dB indicates a scenario in which the signal power equals the noise power.
The scatterer densities range from 0 to 0.2387 scatterers/\umcubic (0 to 6.5723 scatterers/\pixcubic).
Note that here we have randomized the resolutions instead of using the precise resolutions of a specific OCT system. 
This approach was adopted because practical measurements can exhibit varying resolutions due to uncontrolled factors such as defocus, aberrations, and etc.

The CNN consists of three convolutional and max-pooling layer pairs followed by two fully connected layers; this structure is identical to that described in Section 2.2.2 of Ref.\@ \cite{seesan2022}. 
The model was trained to minimize the mean squared error (MSE) between the ground-truth scatterer density values and the network outputs by using the Adam optimizer \cite{Adam_2017} at a learning rate of 10$^{-4}$. 
The batch size of the training was set at 32.
The Input image was represented on a linear scale (i.e., not dB-scale) and was normalized into a [0, 1] range before being input into the model.
The validation set used for the training consists of 100 2D speckle patterns that were extracted from 100 3D speckle patterns; these patterns were generated independently from the training dataset.

\subsection{Evaluation method}
The performances of the SDE were validated both numerically and experimentally.
The numerical validation used numerically generated OCT speckle patterns, whereas the experimental validations used two types of scattering phantom (comprising Intralipid phantoms and microsphere phantoms) and \invitro tumor spheroid samples.
For comparison, we also evaluated our previous SDE \cite{seesan2022}, which is identical to the new SDE with the exception of the noise model used to generate the training dataset.
The process details are described in the following sections.

\subsubsection{Numerical validation}
For the numerical validation, we used 100 2D OCT speckle patterns that were generated using the same method that was used to generate the training dataset.
Each of these 100 2D speckle patterns was extracted from different 3D speckle patterns.
Therefore, all the 2D speckle patterns are independent of each other and are based on different resolutions, SNR values, and scatterer densities.
Note that the noise model used for this generation process is the physically accurate noise model that was described in Section \ref{sec:noise_model}.
Because we know the true density, we can determine the estimation accuracy directly via this numerical validation process.

\subsubsection{Validation using scattering phantoms}
For the phantom-based experimental validations, we used two scattering phantom types, i.e., Intralipid phantoms and microsphere phantoms.

The Intralipid phantoms are composed of Intralipid solutions with various concentrations of 1\%, 2\%, 4\%, 6\%, 8\%, and 10\% (v/v concentration). 
These Intralipid solutions are made from a 20\% stock solution of Intralipid (IL-20, Sigma-Aldrich I141). 
Three phantoms were fabricated for each concentration, giving a total of 18 phantoms.
Note that these phantoms are identical to the phantoms used in our previous study \cite{seesan2022}, and the same measurement datasets that were used in the previous study were also reused in this study.

The microsphere phantoms are formed using a mixture of polystyrene microspheres and agarose gel. 
When compared with the Intralipid droplets, the polystyrene microspheres have well controlled and known sizes and refractive index values, and thus these microspheres are frequently used as phantoms in the biomedical optics field \cite{pogue_review_2006}.
In our study, the microsphere-based phantom is particularly useful because we can compute the true scatterer densities from the product-specific particle concentrations and diameter in each case.

Six phantoms with volume concentrations of 0.1\%, 0.2\%, 0.4\%, 0.6\%, 0.8\%, and 1.0\% were prepared by mixing 1-\um diameter polystyrene microspheres (89904-10ML-F, Sigma-Aldrich) with the agarose gel (A1296-100G, Sigma-Aldrich). 
These concentrations corresponded to scatterer densities of 0.0191, 0.0382, 0.0764, 0.1146, 0.1528, and 0.1910 scatterers/\umcubic, respectively.
The scatterer density was computed here using
\begin{equation}
	\label{eq:ScattererDensity}
	\mathrm{Scatterer\,\,\,density} = \frac{6\sigma}{ \pi d^3},
\end{equation}
where $\sigma$ is the volume concentration of the scattering medium and $d$ is the microsphere diameter.
The mixture was poured into an acrylic container with a thickness of 1 mm.  
The container was then refrigerated for 1 h at 5 \textcelsius. 
Note that these phantoms were fabricated by following the protocol described in \cite{mustari_agarose-based_2018}.
Three phantoms were made with each concentration, and thus 18 phantoms were fabricated in total.

Both the Intralipid phantoms and the microsphere phantoms were measured using a swept-source OCT with a probe beam wavelength of 1.3 \um.
(The OCT system is described in detail in Section \ref{sec:OCT}.)
For the measurements, we intentionally attenuated the probe beam by applying a variable neutral density (ND) filter to alter the SNR.
Each phantom was measured with three different round-trip attenuations, which were 0, -5.4, and -11.8 dB for the Intralipid phantoms, and 0, -5.6, and -12.0 dB for the microsphere phantoms.
The probe power on the sample was 12 mW with 0-dB attenuation.

\subsubsection{Human breast cancer spheroid}
Tumor spheroids composed of human breast adenocarcinoma (MCF-7 cell line) were used to evaluate the SDEs from a biomedical application perspective. 
The cells were cultured for 15 days and spheroids with a size of approximately 500 \um were formed. 
The scatterers within the cells are believed to consist of cell nuclei and organelles.
For the measurements, each spheroid was extracted from the culturing environment and then placed in a room-temperature culture medium without CO$_2$ supply.

We performed two types of measurements.
The first type involves longitudinal hourly measurements for up to 28 h, where the measurements were performed using a 2D cross-sectional scan protocol.
Each cross-sectional image (i.e., B-scan) consists of 512 A-lines.
Note that this experiment was performed initially for the study described in Ref.\@ \cite{abd_el-sadek_optical_2020}, and that the same data set was used in our previous SDE study \cite{seesan2022}.

The second type is a 3D measurement taken at two longitudinal time points of 0 h and 20 h.
A 3D datasets consists of 512 $\times$ 128 A-lines.
This measurement was originally performed for the study detailed in Ref. \cite{el-sadek_three-dimensional_2021}.

\subsubsection{Statistical analysis}
For the numerical validation and the polystyrene microsphere phantom experiment, the agreement between the estimates and the ground truth was evaluated statistically via intraclass correlation (ICC) \cite{Shrout1979PhycBull}. 
ICC is a statistical measure employed to evaluate the level of agreement between two datasets, typically denoted as $X$ and $Y$.
It assesses the extent to which these datasets conform to the $x = y$ line, examining both their linear relationship and equality.
And a higher (i.e., closer to 1.0) ICC indicates a better estimate.
This statistical analysis was performed using the \texttt{intraclass\_corr(\quad)} function in the statistics library of Python (Pingouin 0.5.3) in Python 3.7.

\subsubsection{OCT devices and measurement protocol}
\label{sec:OCT}
A polarization-sensitive Jones matrix swept-source OCT system \cite{li_2017, miyazawa_polarization-sensitive_2019} was used to perform the experimental validations.
The probe wavelength was 1.3 \um and the measurement speed was 50,000 A-lines/s.
The axial optical resolution and the pixel separation were 14.1 \um and 7.24 \um (both in the tissue), respectively, while the lateral optical resolution and the pixel separation were 18.1 \um and 1.95 \um, respectively. 
Note that the pixel separations were identical to those of the numerically generated speckle patterns.
The OCT image used in this study is a coherent composition of multiple polarization channels that is nearly identical to a standard non-polarization-sensitive OCT image.

\section{Results}
\label{sec:results}

\subsection{Numerical validation}
\label{sec:res-numerical}
\begin{figure}
	\centering
	\includegraphics[width=3in]{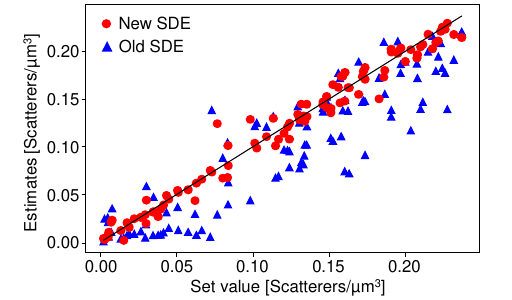}
	\caption{%
		Numerical validation of the new SDE (red circles) and the old SDE (blue triangles).
		The horizontal and vertical axes correspond to the set scatterer density (the ground truth) and the estimated densities, respectively.
		The black solid lines represents the perfect estimate.
		The new SDE gives reasonable estimates, whereas the estimates from the old SDE are downshifted significantly from the ground truth.
		}
	\label{fig:Numer_SD}
\end{figure}
Figure \ref{fig:Numer_SD} shows the numerical validation results, where the estimated scatterer densities are plotted versus the set (ground truth) scatterer densities.
The SDE that was trained using the physically accurate noise model (the new SDE, red circles) shows high consistency between the set and estimated scatterer densities. 
In contrast, the previous SDE \cite{seesan2022}, which was trained using a spatially-uncorrelated noise model (the old SDE, blue triangles), shows higher variation among the estimates, and the estimated values are also downshifted significantly from the ground truth.
The ICCs between the estimates and the ground truth were computed to be 0.975 for the new SDE and 0.722 for the old SDE.
The higher ICC of the new SDE is a quantitative demonstration of the superior performance of the new SDE.

\subsection{Intralipid phantom}
\label{sec:res-IL}
\begin{figure}
	\centering
	\includegraphics[width=3in]{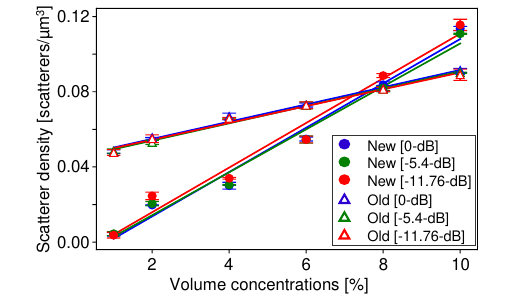}
	\caption{
		Scatterer density estimation results for Intralipid phantoms with several volume concentrations and several probe-beam attenuations.
		The circles and the triangles represent the estimates obtained when using the new and old SDEs, respectively.
		The colors represent the probe beam attenuations that correspond to the SNR of the OCT image.
		The error bar shows the standard deviation for three measurements of the three phantoms.
		Both SDEs show high repeatability (i.e., a small standard deviation in each case) and consistent estimates among the different probe-beam attenuations. 
		However, the old SDE shows a significantly large intercept that indicates the low fidelity of the old SDE, particularly at low concentrations.
		The lines are linear regression lines computed from the data.
	}
	\label{fig:Intralipid}
\end{figure}
The results of the Intralipid phantom validation are summarized in Fig.\@ \ref{fig:Intralipid}.
In this figure, the estimated scatterer densities are plotted versus the volume concentrations.
The circles and triangles represent the estimates from the new and old SDEs, respectively.
The colors of the plots indicate the probe-beam attenuations in each case.
Specifically, each color corresponds to a different SNR.
It should be noted that each cross-sectional OCT B-scan is processed by each estimator using a sliding kernel with a size of 16 $\times$ 16 pixels to obtain a scatterer density image. 
The mean scatterer density of the scattering phantom region is then computed. 
Since there are three samples, three mean scatterer densities were obtained under each measurement condition. 
These three mean scatterer densities were further averaged and plotted. 
The error bars represent the standard deviations among the three measurements of the three phantoms and the lines are linear regression lines that were computed from the data.

Both SDEs showed reasonably small variations (standard deviations) among the phantoms, and shoewd highly linear relationships between the estimates (on the vertical axis) and the volume concentrations (on the horizontal axis).
In addition, the estimated values were not sensitive to probe attenuation.
Specifically, both SDEs were unaffected by the SNR of the OCT image.
However, the intercept of the old SDE was as high as 0.05 \uminvcubic at the 0\% volume concentration, whereas that of the new SDE was close to zero 
(0.01 \uminvcubic).
Because the scatterer density at the 0\% volume concentration might be 0 \uminvcubic, we can conclude that the new SDE provides significantly better estimation accuracy when compared with the old SDE.

\subsection{Polystyrene microsphere phantom}
\label{sec:res-ms}
\begin{figure}
	\centering
	\includegraphics[width=3in]{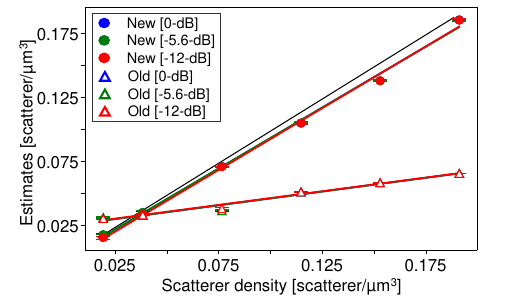}
	\caption{
		Scatterer density estimation results for the microsphere phantoms.
		The vertical axis represents the estimated scatterer density and the horizontal axis represents the theoretically predicted scatterer density as computed from the volume concentration of the phantom using Eq.\@ (\ref{eq:ScattererDensity}).
		The error bars represents the standard deviations among the three estimates of the three samples. 
		The black line represents the ideal estimate, while the other lines are linear-regression lines that were computed from the data.
		The error bars are almost unrecognizable because of the high repeatability of the estimates.
		The new SDE (circles) provides estimates that are very close to the theoretical predictions (black line).
		In contrast, the estimate of the old SDE show significant departures from the theoretical predictions.
		In addition, the estimation results are unaffected by the probe-beam attenuation.
		Specifically, although the attenuations are represented by different colors, the plots and the regression lines with the different attenuations overlap significantly and are not really distinguishable.
	}
	\label{fig:Microsprere}
\end{figure}
The estimation results for the microsphere phantoms are plotted versus the theoretically computed scatterer density in Fig.\@ \ref{fig:Microsprere}.
Here, the plotted values represent the average scattering densities among three phantoms, obtained using the same method as that employed for the Intralipid phantom (refer to the first paragraph of Section \ref{sec:res-IL} for details). 
The theoretical scatterer density was calculated from the volume concentration of the phantom using Eq.\@ (\ref{eq:ScattererDensity}).
Similar to Fig.\@ \ref{fig:Intralipid}, the circles and the triangles represent the estimates from the new and old SDEs, respectively, and the colors indicate the probe-beam attenuation levels.
The error bars indicate the standard deviations among the three measurements of the three phantoms, but the error bars are nearly invisible here because of the very high measurement repeatability (leading to very small standard deviations).
In addition, both SDEs are nearly perfectly independent of the SNRs, as indicated by the fact that the data points for the different SNRs are nearly perfectly overlapping each other.

The estimates from both SDEs show highly linear relationships with the theoretically predicted scatterer density.
However, the estimates from the old SDE (triangles) show significant departures from the ideal estimation line (i.e., the solid black line).
In contrast, the new SDE (circles) shows very close agreement with the theoretical predictions.
The ICCs between the estimateds values and the theoretical scatterer density values were very high at 0.982 (0-dB attenuation), 0.982 (-5.6-dB attenuation), and 0.981 (-12-dB attenuation) for the new SDE, whereas those for the old SDE were 0.467 (0-dB attenuation), 0.467 (-5.6-dB attenuation), and 0.460 (-11.8-dB attenuation).
The higher ICCs indicate the superior performance of the new SDE.

\subsection{Tumor spheroids}
\label{sec:res-spheroid}
\begin{figure}
	\centering
	\includegraphics{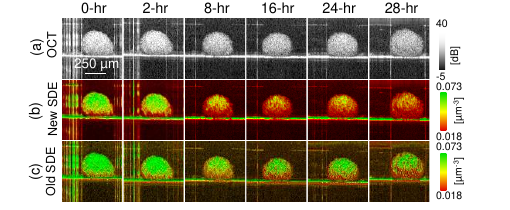}
	\caption{
		Time-lapse images of a tumor spheroid. The individual rows show (a) the OCT intensity images, and the scatterer density images obtained by the (b) new and (c) old SDEs.
		Both SDEs show a reduction in the scatterer density over time.
		}
	\label{fig:Spheroid}
\end{figure}
\begin{figure}
	\centering
	\includegraphics[width=3in]{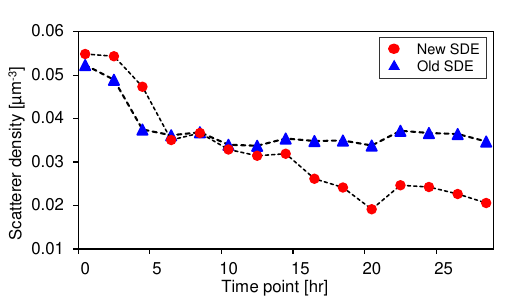}
	\caption{
		Mean scatterer density within the spheroid region.
		Both the new (red circles) and old (blue rectangles) SDEs showed a reduction in the mean scatterer density over time.
		For the time points after 10 h, the old SDE gives constant scatterer density estimates, whereas the new SDE shows a continuous reduction in the scatterer density. 
		As discussed in Section \ref{sec:reductionInSpheroid}, the results from the new SDE are more plausible in this case.
		}
	\label{fig:SD-time-course-average}
\end{figure}

Figure \ref{fig:Spheroid} shows hourly time-lapse images of the MCF-7 spheroid. 
The individual rows show the OCT B-scans, the scatterer density images obtained using the new SDE, and those obtained using the old SDE (in order from top to bottom).
The color scale (hue) in the scatterer density images represents the estimated scatterer density, while the brightness corresponds to the OCT intensity. 
Note that the images from the old SDE are identical to those presented in Ref.\@ \cite{seesan2022}.
In addition, the mean scatterer density was computed within manually segmented spheroid regions and plotted versus time, as shown in Fig.\@ \ref{fig:SD-time-course-average}.
In the figure, the red circles and the blue rectangles correspond to the new and old SDEs, respectively.

Both SDEs showed a reduction in the scatterer density over time, but distinctive differences between the two SDEs can be seen at the later time points after 10 h.
At these time points, the new SDE showed a continuous reduction in the scatterer density, whereas the estimates from the old SDE remained constant.
The constant estimation from the old SDE may be caused by the relatively large error of low SDE at lower scatterer densities, as shown in the numerical and phantom-based validations presented in Sections \ref{sec:res-numerical}, \ref{sec:res-IL}, and \ref{sec:res-ms}.
Here, we remind that the low scatterer density may cause a lower OCT signal intensity and thus lead to the higher noise dominance.
Therefore, the scatterer density estimation can be more helpful in maintaining the noise model accuracy for the lower scatter density cases.
The continuous reduction in the scatterer density found by the new SDE is more consistent with to our dynamic OCT findings than the results from the old SDE; this will be addressed later in the discussion section (Section \ref{sec:reductionInSpheroid}).

\begin{figure}
	\centering
	\includegraphics{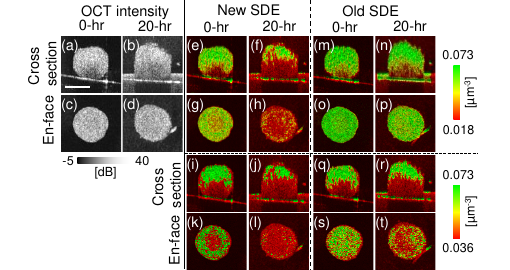}
	\caption{
		3D time-course images of MCF-7 spheroid.
		(a-d) show OCT intensity images, and (e-l) and (m-t) are scatterer density images acquired when using the new and old estimators, respectively.
		The first and third rows represent the cross-sectional images, while the second and fourth rows represent \enface images, respectively.
		At the zero-hour time point, the OCT intensity did not show a clear structure within the spheroid, whereas the \enface scatterer density images showed a clear low-density core at their center. 
		In addition, the \enface scatterer density images showed clear difference between two time points, while this stark contrast was not observed in the \enface OCT intensity images. 
		The scale bar represents 200 \um.}
	\label{fig:3DSpheroid}
\end{figure}
Figure \ref{fig:3DSpheroid} shows cross-sectional (first and third rows) and \enface (second and fourth rows) images obtained from a volumetric dataset of a tumor spheroid. 
The \enface images are the slices acquired around the equator of the spheroid.
The left images (a-d) are the OCT intensity images, the middle images (e-l) are the scatterer density images acquired using the new SDE, and the right (m-t) are those acquired when using the old SDE.
Each scatterer density image is shown with two colormaps since the optimal display range for new- and old-SDE images are not the same. 
And hence, we show each image with two different colormaps. 
Namely, the first and third rows show the cross-sectional images in two different colormaps, similarly the second and fourth rows show \enface images in the two colormaps.

The cross-sectional images showed similar appearances to the corresponding images in Fig.\@ \ref{fig:Spheroid}.
However, the \enface images illustrate that the scatterer density images are more informative than the OCT intensity images.
For example, the OCT intensity at zero-hour time point [Fig.\@ \ref{fig:3DSpheroid}(c)] only shows a homogeneous appearance.
In contrast, the scatterer density image from the new SDE [Fig.\@ \ref{fig:3DSpheroid}(g, k)] revealed that the center region has a lower scatterer density than the peripheral regions. 
This may be an indication of the well-known necrotic core of MCF-7 spheroids \cite{Hirschhaeuser2010JBT, Costa2016BA, el-sadek_three-dimensional_2021}.
Notably, the core is not visible when using the old SDE [Fig.\@ \ref{fig:3DSpheroid}(o, s)].
This may occur because the estimator accuracy of the old SDE is too low for the low scatterer density in this case.

The \enface OCT intensity images at the two time points [Fig.\@ \ref{fig:3DSpheroid}(c) and (d)] do not show clear differences, whereas the scatterer density images [Fig.\@ \ref{fig:3DSpheroid}(g, k) versus (h, i) and Fig.\@ \ref{fig:3DSpheroid}(o, s) versus (p, t)] show clear reductions in the scatterer density at the late time point (20 h).
The reduction in the scatterer density at the later time point is reasonable because the dominant scatterers in these cells are the nuclei and cell organelles, and they are resolved by cell death.
Some discussions related to this point can be found in Section \ref{sec:reductionInSpheroid}.
It should also be noted that the scatterer density from the old SDE at the later time point [Fig.\@ \ref{fig:3DSpheroid}(p, t)] is higher than that obtained from the new SDE [Fig.\@ \ref{fig:3DSpheroid}(h, l)]. 
This difference also can be attributed to the low accuracy of the old SDE for low scatterer densities.

Additionally, we can see that the scatterer densities of the upper and lower parts of the spheroid are not symmetrical. 
Specifically, the upper hemisphere shows a higher scatterer density than the lower hemisphere. 
This issue is discussed in greater detail in Section \ref{sec:Asymetric_spheroid}.

\section{Discussion}
\subsection{Time course reduction and scatterer density}
\label{sec:reductionInSpheroid}
Both the 2D and 3D spheroid imaging results presentrd in Section \ref{sec:res-spheroid} showed time-course reductions in the scatterer densities.
As shown in Fig.\@ \ref{fig:SD-time-course-average}, the new SDE indicates a continuous reduction in the scatterer density over 28 h, while the scatterer density estimated using the old SDE becomes constant after 10 hours.

These reductions in the scatterer density can be understood more easily when using dynamic OCT imaging.
In our previous paper, the same datasets used in Figs.\@ \ref{fig:Spheroid} and \ref{fig:SD-time-course-average} were analyzed using two dynamic OCT methods, i.e., the logarithmic intensity variance (LIV) method and the late OCT correlation decay speed (OCDS$_l$) method.
The mean LIV and mean OCDS$\l$ values within the entire spheroid region were found to be decreasing continuously over 28 hours.
These reductions in the dynamic OCT signals indicate the progression of necrosis within the spheroid \cite{el-sadek_three-dimensional_2021}.

The necrotic cell death causes destruction of the nuclei and the cell organelles.
Because the nuclei and the organelles are the dominant scatterers within the cell, the progression of the necrosis causes a reduction in the scatterer density.

According to the dynamic OCT analysis, the necrosis has progressed continuously over 28 h; this may also suggest that the reduction in the scatterer density also may progress over this time period.
Therefore, the results from the new SDE, which showed a continuous reduction in the scatterer density, are more plausible than the results from the old SDE, in which the scatterer density becomes constant after 10 h.

\subsection{Asymmetric appearance of the scatterer density in spheroids}
\label{sec:Asymetric_spheroid}
Cross-sectional scatterer density images of the spheroid, where Fig.\@ \ref{fig:Spheroid} (b) and (c) show asymmetric appearances between the upper and lower parts of the spheroid.
This asymmetry may partially be attributed to the occurrence of multiple scattering in the deeper region of the tissue. 
Because our speckle generator does not incorporate the multiple scattering effect, the estimation accuracy of the SDE may decrease in the deeper region.

Several possible solutions have been proposed to address this issue.
One approach is to apply one of the available multiple-scattering rejection methods \cite{borycki_spatiotemporal_2019, auksorius_crosstalk-free_2019, auksorius_multimode_2022, LidaZhu2023BOEa}. 
Another possible solution is to improve the speckle generator to allow it to account for the multiple scattering effect. 
Several models and simulation methods have been proposed that can account for multiple scattering \cite{Schmitt1997JOSAA, Thrane2000JOSAA, karamata_multiple_2005, Munro2015OpEx, Munro2016OpEx}.
Although these current methods are computationally intensive and rather too slow to generate the massive amount of training data required, future investigations of the theoretical model and associated simulation methods may enable development of a speckle generator that can account for multiple scattering with reasonable computation speeds. 

\subsection{The impact of noise types on the estimation accuracy}
\begin{figure}
	\centering
	\includegraphics{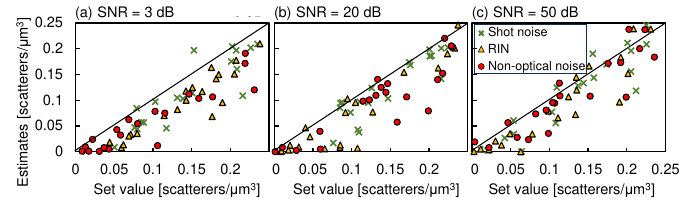}
	\caption{
		Numerical validation of the new SDE was conducted for each noise type.
		In the plot, the color and shape of the points represent the noise type, while the black line indicates the ideal estimation.
		It was observed that the estimates approach the ideal line as the SNR increases. However, no significant dependency on noise type was noted.}
	\label{fig:constantNoise}
\end{figure}
Since we distinguished three types of the noise to train the new estimator, it was anticipated that the estimation accuracy may depend on the dominant noise type of the input image. 
This issue was investigated by using numerically generated speckle patterns, and it was found that the estimation accuracy was not significantly affected by the dominant noise types. 
The details of this study are as follows.

A total of 180 3D speckle patterns were generated, each with resolutions and scatterer density randomly selected from the same ranges as those used in the training-set generation (Section \ref{sec:speckleGenerator}).
From each of these 180 3D speckle patterns, one 2D speckle pattern was extracted, resulting in 180 fully independent 2D speckle patterns. 
To each 2D speckle pattern, only one type of noise, i.e., shot-noise, RIN, or non-optical noise, was added. 
The SNR was set to 3, 20, or 50 dB. 
Finally, 20 noisy 2D speckle patterns were generated for each combination of SNR and noise type.

The speckle images were processed by the new SDE, and the estimation results are summarized in Fig.\@ \ref{fig:constantNoise}. 
In this plot, each color and shape of the plot points indicate the type of noise, while the black line represents the ideal estimation line. 
It was found that the estimates are closer to the ideal line with higher SNR for all noise types. 
However, no marked dependency on noise type was observed across all SNRs.

\subsection{Open issue: Physical mechanism used in the SDE}
\begin{figure}
	\centering
	\includegraphics{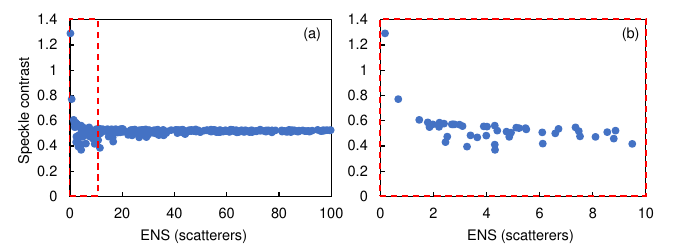}
	\caption{
		The simulated speckle contrasts as a function of ENS, where the speckle patterns were numerically generated with the same resolution- and scatterer-density ranges with the validation dataset of Section \ref{sec:res-numerical}.
		The speckle contrast becomes almost constant if the ENS is larger than a few scatterers.
		(b) provides a zoomed-up view of the red box in (a).}
	\label{fig:SpeckleContrast-ENS}
\end{figure}
As we have described in the introduction, the development of the improved noise model was motivated by our anticipation about the possible working principle of the CNN-based SDE. 
Specifically, we suspected that the SDE might internally estimate speckle contrast and resolutions from speckle patterns. 
Speckle contrast was expected to aid in estimating the number of scatterers in the resolution volume (ENS), while resolution volume could be derived from estimated resolutions (as discussed in Section 4.3 of Ref.\@ \cite{seesan2022}). 
This principle could work effectively if speckle contrast were sensitive to ENS, particularly for small ENS values

However, we found that our SDE could function not only in this regime but also for cases with large ENS values. 
For instance, Fig.\@ \ref{fig:SpeckleContrast-ENS} depicts the speckle contrast of OCT speckles as a function of ENS, where speckles were numerically generated with the same parameter ranges of resolutions and scatterer density as the validation dataset used in Section \ref{sec:res-numerical}, and noise was added to the speckle pattern.
Speckle contrast is defined as the standard deviation of signal amplitude over the 3D speckle pattern divided by the mean amplitude. 
The ENS is defined in the same manner as described in Ref.\@ \cite{hillman_correlation_2006}, where it represents the number of scatterers within a volume of $\Delta x$ $\times$ $\Delta y$ $\times$ $\Delta z$. 
Here, $\Delta x$ and $\Delta y$ denote the lateral resolutions, defined as the $1/e^2$-width of the intensity, while $\Delta z$ represents the axial resolution, defined as the FWHM of the intensity. 
As illustrated in this plot, the speckle contrast remains almost constant across most of the ENS ranges.

During our numerical validation and phantom experiments, the new SDE provided accurate estimates even for ENS values around a few hundred. 
Namely, the SDE worked even for high scatterer density samples where the speckle contrast remains nearly constant.

We suspect that the SDE estimator leverages additional properties of speckle beyond speckle contrast and resolutions. 
One such property might be the spatial shape or statistics of the speckles. 
Given that this study demonstrated improved estimation accuracy by considering the spatial properties of noise and signal, this suspicion appears plausible.

Our results affirm the SDE's strong estimation performance, yet further analysis of the trained model and investigation into potentially exploited physical principles remain important.

\subsection{Computation time of the speckle generator and the SDE}
The overall training process of the CNN-based SDE can be split into two sub-processes, i.e., speckle generation and CNN training.
The generation of 80,000 3D speckle patterns with dimensions of 16 $\times$ 16 $\times$ 16 pixels takes approximately 3 hours when using a desktop PC equipped with an Intel Core i7-6900K CPU and a graphics processing unit (GPU; GeForce RTX 3090 Ti, NVIDIA).
The main memory of the PC is 128 GB in size, and the GPU has 10752 cores, a 1.86 GHz boost clock, and 24 GB of memory.
The speckle generator is written in Python (version 3.7) with a GPU-compatible numerical computation library (CuPy 12.0.0).
Notably, the speckle generator is more than 11 times faster than our previous version, which did not use a GPU.
Therefore, usage of a GPU is essential when generating large training datasets.

Subsequently, training of the CNN model takes approximately 1 h, where the CNN is written in Python 3.7 with Keras 2.3.1 based on the TensorFlow backend.
Therefore, the entire training process takes approximately 4 hours.

Scatterer density estimation (i.e., inference) of a 512 $\times$ 402 pixel cross-sectional image using a 16 $\times$ 16 pixels sliding window took around 1 minute, and thus the estimation process for a volumetric image that consists of 128 B-scans takes approximately 2 hours.
The inference can be faster by using more optimized network architecture and/or implementation, and this would be a future research topic.

\subsection{Compatibility of the trained model and the scanning protocol}
Because the speckle generator generates a speckle pattern with a specific pixel size, each training dataset (i.e., each set of speckle patterns) is specific to a single scan protocol.
Therefore, speckle patterns should be generated for each different scan protocol and a specific SDE model should then be trained using these specific speckle patterns. 

The other compatibility issue that must be considered is the minimum separation required between adjacent A-lines (i.e., the lateral pixel size).
We suspect that our new SDE uses the different lateral spatial properties of the noise and OCT signals to distinguish them, as summarized in Table \ref{tab:spatialCorrelation}.
Specifically, we have assumed that the noise has fully decorrelated between the adjacent A-lines, as described in Section \ref{sec:noise_model}.
However, the OCT signals are mutually correlated along the lateral direction with the correlation distance around the lateral resolution.
The different lateral spatial properties of the noise and the OCT may have been used by the SDE model to discriminate the OCT signal from the noise.
However, the different lateral extents only become pronounced if the A-line separation is smaller than the lateral resolution.
Therefore, the OCT scan protocol to be used with our SDE should have an A-line separation that is smaller than the lateral resolution.

It is also notable that, according to Eq.\@ (\ref{eq:Na_zDomain}), the RIN may have the same axial extent as the OCT signal.
Thus, if the CNN model cannot use the different lateral properties of the signal and the noise, as in the case where the A-line separation is far greater than the lateral resolution, the CNN model cannot discriminate the RIN from the OCT signal.
This further emphasizes the need to consider the minimum separation required for the A-lines.

\section{Conclusion}
In this study, we demonstrated a CNN-based SDE that is trained using speckle patterns generated by a numerical speckle generator.
This speckle generator uses a physically realistic noise model that takes the spatial properties of shot noise, RIN, and non-optical noise into account. 
The SDE was examined using numerically generated OCT speckle patterns along with the experimentally obtained OCT images of two types of scattering phantom and \invitro tumor spheroid samples.
In these examinations, the SDE's estimation performance was compared with that of our previous CNN-based SDE, which was trained with speckle patterns that did not account for the spatial properties of the noise.
All the validation experiments demonstrated the superior estimation accuracy provided by the new SDE.
In particular, the validation using microsphere phantoms demonstrated the excellent agreement of the estimated scatterer density with the ground truth. 

In future work, this new SDE can be used in a variety of biomedical application to assess the sub-resolution structural properties of tissues and cells.


\section*{Disclosures}
T. Seesan, P. Mukherjee, I.A. El-Sadek, Y. Lim, L. Zhu, S. Makita, Y. Yasuno: Sky Technology (F), Nikon (F), Kao Corp. (F), Topcon (F). 

\section*{Funding}
Core Research for Evolutional Science and Technology (JPMJCR2105); 
Japan Society for the Promotion of Science  (21H01836, 22K04962, 22KF0058); 
Japan Science and Technology Agency (JPMJFS2106);
China Scholarship Council (201908130130).

\section* {Data, Materials, and Code Availability} 
Data underlying the results presented in this paper are not publicly available at this time but may be obtained from the authors upon reasonable request. 

\bibliography{References}

\begin{thebibliography}{10}
\newcommand{\enquote}[1]{``#1''}

\bibitem{huang_optical_1991}
D.~Huang, E.~A. Swanson, C.~P. Lin, J.~S. Schuman, W.~G. Stinson, W.~Chang, M.~R. Hee, T.~Flotte, K.~Gregory, C.~A. Puliafito, and J.~G. Fujimoto, \enquote{Optical coherence tomography,} {\protect\JournalTitle{Science}} \textbf{254}, 1178--1181 (1991).

\bibitem{fujimoto_optical_2000}
J.~G. Fujimoto, C.~Pitris, S.~A. Boppart, and M.~E. Brezinski, \enquote{Optical coherence tomography: an emerging technology for biomedical imaging and optical biopsy,} {\protect\JournalTitle{Neoplasia}} \textbf{2}, 9--25 (2000).

\bibitem{fujimoto_optical_2003}
J.~G. Fujimoto, \enquote{Optical coherence tomography for ultrahigh resolution in vivo imaging,} {\protect\JournalTitle{Nat. Biotechnol.}} \textbf{21}, 1361--1367 (2003).

\bibitem{Chen2007}
J.~Chen and L.~Lee, \enquote{Clinical applications and new developments of optical coherence tomography: an evidence‐based review,} {\protect\JournalTitle{Clin. Exp. Optom.}} \textbf{90}, 317--335 (2007). PMID: 17697178.

\bibitem{liu_tissue_2017}
S.~Liu, \enquote{Tissue characterization with depth-resolved attenuation coefficient and backscatter term in intravascular optical coherence tomography images,} {\protect\JournalTitle{J. Biomed. Opt.}} \textbf{22}, 1 (2017).

\bibitem{chang_review_2019}
S.~Chang and A.~K. Bowden, \enquote{Review of methods and applications of attenuation coefficient measurements with optical coherence tomography,} {\protect\JournalTitle{J. Biomed. Opt.}} \textbf{24}, 1 (2019).

\bibitem{huang_optical_2017}
Y.~Huang, S.~Wang, Q.~Guo, S.~Kessel, I.~Rubinoff, L.~L.-Y. Chan, P.~Li, Y.~Liu, J.~Qiu, and C.~Zhou, \enquote{Optical coherence tomography detects necrotic regions and volumetrically quantifies multicellular tumor spheroids,} {\protect\JournalTitle{Cancer Res.}} \textbf{77}, 6011--6020 (2017).

\bibitem{huang_longitudinal_2020}
Y.~Huang, J.~Zou, M.~Badar, J.~Liu, W.~Shi, S.~Wang, Q.~Guo, X.~Wang, S.~Kessel, L.~L.-Y. Chan, P.~Li, Y.~Liu, J.~Qiu, and C.~Zhou, \enquote{Longitudinal morphological and physiological monitoring of three-dimensional tumor spheroids using optical coherence tomography,} {\protect\JournalTitle{J. Vis. Exp.}}  (2020).

\bibitem{cauberg_quantitative_2010}
E.~C.~C. Cauberg, D.~M. de~Bruin, D.~J. Faber, T.~M. de~Reijke, M.~Visser, J.~J. M. C.~H. de~la Rosette, and T.~G. van Leeuwen, \enquote{Quantitative measurement of attenuation coefficients of bladder biopsies using optical coherence tomography for grading urothelial carcinoma of the bladder,} {\protect\JournalTitle{J. Biomed. Opt.}} \textbf{15}, 066013 (2010).

\bibitem{zhao_ex_2012}
Q.~Zhao, C.~Zhou, H.~Wei, Y.~He, X.~Chai, and Q.~Ren, \enquote{\textit{{Ex} vivo} determination of glucose permeability and optical attenuation coefficient in normal and adenomatous human colon tissues using spectral domain optical coherence tomography,} {\protect\JournalTitle{J. Biomed. Opt.}} \textbf{17}, 1050041 (2012).

\bibitem{muller_prostate_2016}
B.~G. Muller, D.~M. de~Bruin, M.~J. Brandt, W.~van~den Bos, S.~van Huystee, D.~J. Faber, D.~Savci, P.~J. Zondervan, T.~M. de~Reijke, M.~P. Laguna-Pes, T.~G. van Leeuwen, and J.~J. M. C.~H. de~la Rosette, \enquote{Prostate cancer diagnosis by optical coherence tomography: {First} results from a needle based optical platform for tissue sampling,} {\protect\JournalTitle{J. Biophotonics}} \textbf{9}, 490--498 (2016).

\bibitem{van_leeuwen_measurement_2003}
T.~G. van Leeuwen, D.~J. Faber, and M.~C. Aalders, \enquote{Measurement of the axial point spread function in scattering media using single-mode fiber-based optical coherence tomography,} {\protect\JournalTitle{IEEE J. Sel. Top. Quantum Electron.}} \textbf{9}, 227--233 (2003).

\bibitem{stefan_determination_2018}
S.~Stefan, K.-S. Jeong, C.~Polucha, N.~Tapinos, S.~A. Toms, and J.~Lee, \enquote{Determination of confocal profile and curved focal plane for {OCT} mapping of the attenuation coefficient,} {\protect\JournalTitle{Biomed. Opt. Express}} \textbf{9}, 5084 (2018).

\bibitem{li_robust_2020}
K.~Li, W.~Liang, Z.~Yang, Y.~Liang, and S.~Wan, \enquote{Robust, accurate depth-resolved attenuation characterization in optical coherence tomography,} {\protect\JournalTitle{Biomed. Opt. Express}} \textbf{11}, 672 (2020).

\bibitem{baumann_signal_2019}
B.~Baumann, C.~W. Merkle, R.~A. Leitgeb, M.~Augustin, A.~Wartak, M.~Pircher, and C.~K. Hitzenberger, \enquote{Signal averaging improves signal-to-noise in {OCT} images: {But} which approach works best, and when?} {\protect\JournalTitle{Biomed. Opt. Express}} \textbf{10}, 5755--5775 (2019).

\bibitem{seesan_intensity-invariant_2020}
T.~Seesan, D.~Oida, K.~Oikawa, P.~Buranasiri, and Y.~Yasuno, \enquote{Intensity-invariant scatterer density estimation for optical coherence tomography using deep convolutional neural network,} {\protect\JournalTitle{Proc. SPIE, Biomedical Imaging and Sensing Conference 2020}} \textbf{11521}, 115210Q (2020).

\bibitem{seesan_sample_2021}
T.~Seesan, D.~Oida, K.~Oikawa, P.~Buranasiri, and Y.~Yasuno, \enquote{Sample and system parameter estimation from local speckle pattern by fully numerically trained deep convolution neural network,} {\protect\JournalTitle{Proc. SPIE, Optical Coherence Tomography and Coherence Domain Optical Methods in Biomedicine XXV}} \textbf{11630}, 1163021 (2021).

\bibitem{seesan2022}
T.~Seesan, I.~Abd El-Sadek, P.~Mukherjee, L.~Zhu, K.~Oikawa, A.~Miyazawa, L.~T.-W. Shen, S.~Matsusaka, P.~Buranasiri, S.~Makita, and Y.~Yasuno, \enquote{Deep convolutional neural network-based scatterer density and resolution estimators in optical coherence tomography,} {\protect\JournalTitle{Biomed. Opt. Express}} \textbf{13}, 168 (2022).

\bibitem{hillman_correlation_2006}
T.~R. Hillman, S.~G. Adie, V.~Seemann, J.~J. Armstrong, S.~L. Jacques, and D.~D. Sampson, \enquote{Correlation of static speckle with sample properties in optical coherence tomography,} {\protect\JournalTitle{Opt. Lett.}} \textbf{31}, 190 (2006).

\bibitem{kurokawa_-plane_2015}
K.~Kurokawa, S.~Makita, Y.-J. Hong, and Y.~Yasuno, \enquote{In-plane and out-of-plane tissue micro-displacement measurement by correlation coefficients of optical coherence tomography,} {\protect\JournalTitle{Opt. Lett.}} \textbf{40}, 2153 (2015).

\bibitem{Leitgeb2003OpEx}
R.~Leitgeb, C.~K. Hitzenberger, and A.~F. Fercher, \enquote{Performance of fourier domain vs. time domain optical coherence tomography,} {\protect\JournalTitle{Opt. Express}} \textbf{11}, 889--894 (2003).

\bibitem{deBoer2003OL}
J.~F. de~Boer, B.~Cense, B.~H. Park, M.~C. Pierce, G.~J. Tearney, and B.~E. Bouma, \enquote{Improved signal-to-noise ratio in spectral-domain compared with time-domain optical coherence tomography,} {\protect\JournalTitle{Opt. Lett.}} \textbf{28}, 2067--2069 (2003).

\bibitem{Adam_2017}
D.~P. Kingma and J.~Ba, \enquote{Adam: a method for stochastic optimization,} {\protect\JournalTitle{arXiv:1412.6980 [cs]}}  (2017). ArXiv: 1412.6980.

\bibitem{pogue_review_2006}
B.~W. Pogue and M.~S. Patterson, \enquote{Review of tissue simulating phantoms for optical spectroscopy, imaging and dosimetry,} {\protect\JournalTitle{J. Biomed. Opt.}} \textbf{11}, 041102 (2006).

\bibitem{mustari_agarose-based_2018}
A.~Mustari, I.~Nishidate, M.~A. Wares, T.~Maeda, S.~Kawauchi, S.~Sato, M.~Sato, and Y.~Aizu, \enquote{Agarose-based {Tissue} {Mimicking} {Optical} {Phantoms} for {Diffuse} {Reflectance} {Spectroscopy},} {\protect\JournalTitle{J. Visualized Exp.}} p. 57578 (2018).

\bibitem{abd_el-sadek_optical_2020}
I.~Abd El-Sadek, A.~Miyazawa, L.~Tzu-Wei~Shen, S.~Makita, S.~Fukuda, T.~Yamashita, Y.~Oka, P.~Mukherjee, S.~Matsusaka, T.~Oshika, H.~Kano, and Y.~Yasuno, \enquote{Optical coherence tomography-based tissue dynamics imaging for longitudinal and drug response evaluation of tumor spheroids,} {\protect\JournalTitle{Biomed. Opt. Express}} \textbf{11}, 6231 (2020).

\bibitem{el-sadek_three-dimensional_2021}
I.~A. El-Sadek, A.~Miyazawa, L.~T.-W. Shen, S.~Makita, P.~Mukherjee, A.~Lichtenegger, S.~Matsusaka, and Y.~Yasuno, \enquote{Three-dimensional dynamics optical coherence tomography for tumor spheroid evaluation,} {\protect\JournalTitle{Biomed. Opt. Express}} \textbf{12}, 6844 (2021).

\bibitem{Shrout1979PhycBull}
P.~E. Shrout and J.~L. Fleiss, \enquote{Intraclass correlations: Uses in assessing rater reliability.} {\protect\JournalTitle{Psychol. Bull.}} \textbf{86}, 420--428 (1979).

\bibitem{li_2017}
E.~Li, S.~Makita, Y.-J. Hong, D.~Kasaragod, and Y.~Yasuno, \enquote{Three-dimensional multi-contrast imaging of in vivo human skin by {Jones} matrix optical coherence tomography,} {\protect\JournalTitle{Biomed. Opt. Express}} \textbf{8}, 1290 (2017).

\bibitem{miyazawa_polarization-sensitive_2019}
A.~Miyazawa, S.~Makita, E.~Li, K.~Yamazaki, M.~Kobayashi, S.~Sakai, and Y.~Yasuno, \enquote{Polarization-sensitive optical coherence elastography,} {\protect\JournalTitle{Biomed. Opt. Express}} \textbf{10}, 5162 (2019).

\bibitem{Hirschhaeuser2010JBT}
F.~Hirschhaeuser, H.~Menne, C.~Dittfeld, J.~West, W.~Mueller-Klieser, and L.~A. Kunz-Schughart, \enquote{Multicellular tumor spheroids: An underestimated tool is catching up again,} {\protect\JournalTitle{J. Biotechnol.}} \textbf{148}, 3--15 (2010). Organotypic Tissue Culture for Substance Testing.

\bibitem{Costa2016BA}
E.~C. Costa, A.~F. Moreira, D.~{de Melo-Diogo}, V.~M. Gaspar, M.~P. Carvalho, and I.~J. Correia, \enquote{{3D} tumor spheroids: an overview on the tools and techniques used for their analysis,} {\protect\JournalTitle{Biotechnol. Adv.}} \textbf{34}, 1427--1441 (2016).

\bibitem{borycki_spatiotemporal_2019}
D.~Borycki, M.~Hamkało, M.~Nowakowski, M.~Szkulmowski, and M.~Wojtkowski, \enquote{Spatiotemporal optical coherence ({STOC}) manipulation suppresses coherent cross-talk in full-field swept-source optical coherence tomography,} {\protect\JournalTitle{Biomed. Opt. Express}} \textbf{10}, 2032 (2019).

\bibitem{auksorius_crosstalk-free_2019}
E.~Auksorius, D.~Borycki, and M.~Wojtkowski, \enquote{Crosstalk-free volumetric in vivo imaging of a human retina with {Fourier}-domain full-field optical coherence tomography,} {\protect\JournalTitle{Biomed. Opt. Express}} \textbf{10}, 6390 (2019).

\bibitem{auksorius_multimode_2022}
E.~Auksorius, D.~Borycki, P.~Wegrzyn, I.~Žičkienė, K.~Adomavičius, B.~L. Sikorski, and M.~Wojtkowski, \enquote{Multimode fiber as a tool to reduce cross talk in {Fourier}-domain full-field optical coherence tomography,} {\protect\JournalTitle{Opt. Lett.}} \textbf{47}, 838 (2022).

\bibitem{LidaZhu2023BOEa}
L.~Zhu, S.~Makita, J.~Tamaoki, A.~Lichtenegger, Y.~Lim, Y.~Zhu, M.~Kobayashi, and Y.~Yasuno, \enquote{Multi-focus averaging for multiple scattering suppression in optical coherence tomography,} {\protect\JournalTitle{Biomed. Opt. Express}} \textbf{14}, 4828--4844 (2023).

\bibitem{Schmitt1997JOSAA}
J.~M. Schmitt and A.~Kn\"{u}ttel, \enquote{Model of optical coherence tomography of heterogeneous tissue,} {\protect\JournalTitle{J. Opt. Soc. Am. A}} \textbf{14}, 1231--1242 (1997).

\bibitem{Thrane2000JOSAA}
L.~Thrane, H.~T. Yura, and P.~E. Andersen, \enquote{Analysis of optical coherence tomography systems based on the extended {Huygens}--{Fresnel} principle,} {\protect\JournalTitle{J. Opt. Soc. Am. A}} \textbf{17}, 484--490 (2000).

\bibitem{karamata_multiple_2005}
B.~Karamata, P.~Lambelet, M.~Laubscher, M.~Leutenegger, S.~Bourquin, and T.~Lasser, \enquote{Multiple scattering in optical coherence tomography {I} {Investigation} and modeling,} {\protect\JournalTitle{J. Opt. Soc. Am. A}} \textbf{22}, 1369 (2005).

\bibitem{Munro2015OpEx}
P.~R.~T. Munro, A.~Curatolo, and D.~D. Sampson, \enquote{Full wave model of image formation in optical coherence tomography applicable to general samples,} {\protect\JournalTitle{Opt. Express}} \textbf{23}, 2541--2556 (2015).

\bibitem{Munro2016OpEx}
P.~R.~T. Munro, \enquote{Three-dimensional full wave model of image formation in optical coherence tomography,} {\protect\JournalTitle{Opt. Express}} \textbf{24}, 27016--27031 (2016).

\end{thebibliography}

\end{document}